\documentclass[aps,prb,twocolumn,superscriptaddress,showpacs,amsmath]{revtex4}

\usepackage{graphicx,amssymb}

\begin{document}

%

\title{Pressure dependence of the exchange interaction in the dimeric single-molecule magnet
[Mn$_4$O$_3$Cl$_4$(O$_2$CEt)$_3$(py)$_3$]$_2$ from inelastic neutron scattering}

\author{A. Sieber}
\affiliation{Department of Chemistry and Biochemistry, University of Bern, 3012 Bern, Switzerland}

\author{D. Foguet-Albiol}
\affiliation{Department of Chemistry, University of Florida, Gainesville, Florida 32611, USA}

\author{O. Waldmann}
\affiliation{Department of Chemistry and Biochemistry, University of Bern, 3012 Bern, Switzerland}

\author{S. T. Ochsenbein}
\affiliation{Department of Chemistry and Biochemistry, University of Bern, 3012 Bern, Switzerland}

\author{G. Carver}
\affiliation{Department of Chemistry and Biochemistry, University of Bern, 3012 Bern, Switzerland}

\author{H. Mutka}
\affiliation{Institut Laue-Langevin, 6 Rue Jules Horowitz, BP 156, 38042 Grenoble Cedex 9, France}

\author{F. Fernandez-Alonso}
\affiliation{ISIS Facility, CCLRC Rutherford Appleton Laboratory, Chilton, Didcot, Oxfordshire OX11 0QX, UK}

\author{M. Mezouar}
\affiliation{European Synchrotron Radiation Facility, 6 Rue Jules Horowitz, BP 220, 38043 Grenoble Cedex 9,
France}

\author{H. P. Weber}
\affiliation{ACCE, 38960 St Aupre, France, and Laboratoire de Chrystallographie EFPL-FSB-IPMC, BSP-Dorigny,
1015 Lausanne, Switzerland}

\author{G. Christou}
\affiliation{Department of Chemistry, University of Florida, Gainesville, Florida 32611, USA}

\author{H. U. G\"udel}
\affiliation{Department of Chemistry and Biochemistry, University of Bern, 3012 Bern, Switzerland}

\date{\today}

\begin{abstract}
The low-lying magnetic excitations in the dimer of single-molecule magnets (Mn$_4$)$_2$ are studied by
inelastic neutron scattering as a function of hydrostatic pressure. The anisotropy parameters $D$ and
$B_0^4$, which describe each Mn$_4$ subunit, are essentially pressure independent, while the
antiferromagnetic exchange coupling $J$ between the two Mn$_4$ subunits strongly depends on pressure, with an
increase of 42\% at 17~kbar. Additional pressure dependent powder X-ray measurements allow a structural
interpretation of the findings.
\end{abstract}


\maketitle

%

Molecular magnetic clusters have attracted considerable attention recently because of the striking quantum
phenomena observed in these systems.\cite{Chr00,Mn12_Fe8,Wer99,Hil03,Bar04,Wal05} The so-called
single-molecule magnets (SMM), for instance, exhibit magnetization hysteresis and quantum tunneling of
magnetization (QTM) at low temperatures.\cite{Ses93,Fri96,Tho96}

\begin{figure}
\includegraphics[scale=0.75]{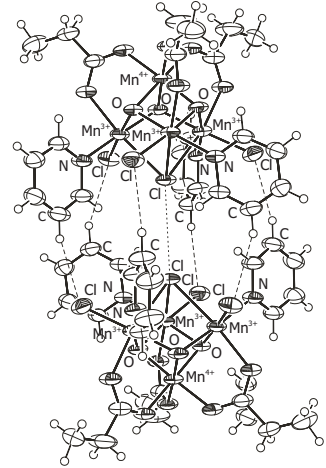}
\caption{\label{fig1} Structure of (Mn$_4$)$_2$. The molecule exhibits a threefold symmetry axis which pases
through the two Mn$^{4+}$ ions. The dashed lines indicate the six C-H$\cdots$Cl hydrogen bonds, and the
dotted line the close Cl$\cdots$Cl approach.}
\end{figure}

An important issue in this context are weak magnetic interactions between individual clusters
(\emph{intermolecular} interactions), as mediated by weak long-distance superexchange paths like hydrogen
bonds. On the one hand, this kind of interactions have to be considered seriously in potential applications
as they would tend to degrade the performance. On the other hand, new physical phenomena can emerge, e.g.,
field-induced long-range ordering in weakly interacting antiferromagnetic (AFM) clusters.\cite{Hos99} Such
phenomena are currently of great interest with regard to magnetization plateaus in quantum spin systems or
Bose-Einstein condensation of magnons.\cite{Nik00,Rue03} Furthermore, intermolecular magnetic interactions
between SMM can provide highly desirable situations, as in the exchange-biased SMM
[Mn$_4$O$_3$Cl$_4$(O$_2$CEt)$_3$(py)$_3$]$_2$, or (Mn$_4$)$_2$. Its dimer structure, shown in
Fig.~\ref{fig1}, consists of two Mn$_4$ monomers which are weakly coupled to each other by AFM exchange
interactions mediated by C-H$\cdots$Cl hydrogen bonds and close Cl$\cdots$Cl contact (see Fig.~\ref{fig1}).
The effect of the weak AFM coupling is striking; it results in a magnetic hysteresis curve with QTM
completely suppressed at zero field,\cite{Wer02} in contrast to the behavior of generic SMM. A precise
understanding of intermolecular interactions between molecular magnetic clusters, SMMs in particular, is thus
important both from the applications and fundamental point of view.

In this work, the pressure dependence of the AFM interaction, of strength $J$, between the Mn$_4$ subunits in
(Mn$_4$)$_2$ is studied by inelastic neutron scattering (INS). The results were complemented with
pressure-dependent powder X-ray measurements. These are the first pressure experiments of a molecular
magnetic compound with weak intermolecular interactions. As a main result a pronounced pressure dependence of
$J$ is found, which correlates to the weak bonds between the Mn$_4$ subunits in (Mn$_4$)$_2$. Our results
further suggest a strong effect of pressure on the magnetic hysteresis curve of (Mn$_4$)$_2$.

%

Partially deuterated microcrystalline samples of (Mn$_4$)$_2$,
[Mn$_4$O$_3$Cl$_4$(O$_2$CEt)$_3$(py-d$_5$)$_3$]$_2$$\cdot$8MeCN, were freshly prepared following
Ref.~\onlinecite{Hen92}, and checked by X-ray powder diffraction and elemental analysis. The INS experiments
were performed on IN5 at the Institut Laue-Langevin (ILL) in Grenoble, France, and on IRIS at the ISIS
Facility, CCLRC Rutherford Appleton Laboratory in Chilton, UK. Spectra were acquired in the temperature range
1.5 - 20~K. On IN5, data were collected for several pressures up to 12~kbar with an incident wavelength
$\lambda_i$ = 7.0~{\AA}. The instrumental resolution was 31~$\mu$eV at the elastic line. On IRIS, data were
recorded at 17~kbar using a pyrolytic graphite (PG002) analyzer with a final wavelength $\lambda_f$ =
6.6~{\AA}. The resolution was 17.5~$\mu$eV at the elastic line. In both experiments the data were corrected
for detector efficiency using a vanadium standard. The spectra correspond to the sum over all detectors.
Further data treatment included subtraction of the background by approximating it with a suitable analytical
function (this function was obtained from the 1.5~K spectrum, which exhibits the least number of magnetic
peaks). For the pressures $p$ = 0 and 5.0(1)~kbar, a standard ILL continuously loaded high-pressure He gas
cell was used (3.7~g sample). For $p$ = 12.0(5)~kbar the standard ILL 15~kbar high-pressure clamp cell (0.4~g
sample), and for $p$ = 17.0(5)~kbar a McWhan high-pressure clamp cell (0.5~g sample) was used, in both cases
with FC-75 (fluorinated hydrocarbon, 3M) as pressure transmitting medium. In all experiments cooling was
achieved with an Orange cryostat.

High-resolution powder diffraction data were collected at ambient temperature and pressures up to 40~kbar on
beam line ID27 at the ESRF, Grenoble. A finely-ground powder sample was inserted into a diamond-anvil cell,
with FC-75 as pressure-transmitting medium, an Inconel gasket to contain the sample, and a ruby crystal as
pressure calibrant. A wavelength of 0.3738~{\AA} was selected with a Si double-crystal monochromator. The
beam was focused to 5~$\mu$m at the sample. The diffracted X-rays were detected with a MAR345 image plate.
Exposure times were of the order of minutes in single-bunch mode. The 2D diffraction data were projected to
1D using the program Fit2D.\cite{Ham} The peak positions were determined with the program FullProf \cite{Rod}
using the profile fitting mode. The space group was determined to R$\bar{3}$. The unit cell parameters $a$
and $c$ were obtained from least-squares fitting the positions of the first ten clearly resolved peaks in the
angular range 1.6$^\circ$ $< 2\theta <$ 4.0$^\circ$.

%

\begin{figure}
\includegraphics[scale=0.75]{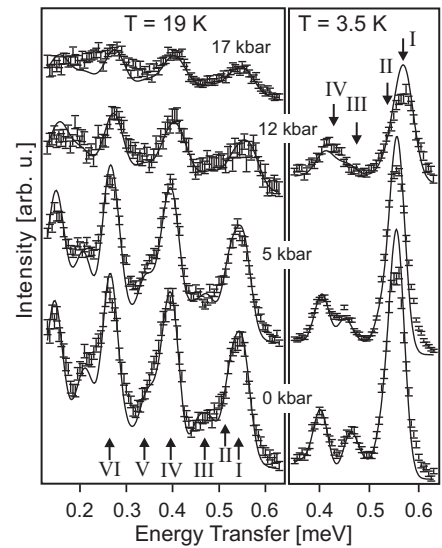}
\caption{\label{fig2} INS spectra of (Mn$_4$)$_2$ recorded on IN5 (data for 0, 5 and 12~kbar at 19 and 3.5~K)
and IRIS (data for 17~kbar at 19~K). Data were background corrected. The labeling of the peaks corresponds to
Fig.~\ref{fig3} (and Ref.~\onlinecite{Sie05}). The solid lines represent the simulated spectra with the
best-fit parameters, and an instrumental resolution of 35~$\mu$eV for 0 and 5~kbar and 45~$\mu$eV for 12 and
17~kbar (the lower resolution at 12 and 17~kbar is attributed to the larger pressure inhomogeneities in clamp
cells as compared to He cells).}
\end{figure}

\begin{figure}
\includegraphics[scale=0.75]{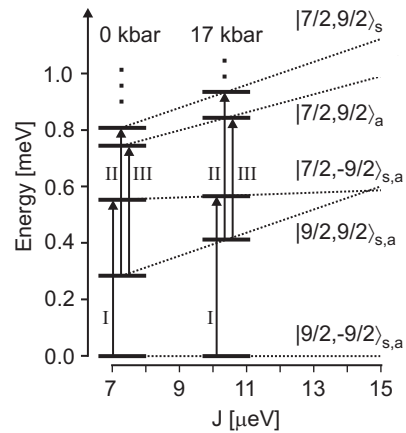}
\caption{\label{fig3} Energy level diagram of (Mn$_4$)$_2$ as function of $J$ ($D$ = -62.4~$\mu$eV, $B^0_4$ =
-7.2$\times$10$^{-3}$~$\mu$eV). The arrows assign the transitions seen in Fig.~\ref{fig2} at 0 and 17~kbar.
The notation for the approximate wave functions is $|M_{S_A},M_{S_B}\rangle_{s/a} = (|M_{S_A},M_{S_B}\rangle
\pm |M_{S_B},M_{S_A}\rangle)/\sqrt{2}$, see Ref.~\onlinecite{Sie05}.}
\end{figure}

Figure~\ref{fig2} presents the neutron energy-loss side of the INS spectra recorded at 19~K and 3.5~K for
pressures from 0 to 17~kbar. Several peaks were observed at transfer energies below 0.6~meV. The 0~kbar data
is fully consistent with our previous ambient pressure INS experiment,\cite{Sie05} and the peaks were labeled
the same way. The broad asymmetric peak at ca. 0.55~meV in fact consists of two close transitions, peak I,
which is the only cold transition, and peak II, which at 3.5~K already has significant intensity. At 3.5~K
peaks III and IV also have significant intensities, and at higher temperatures many more transitions show up
at lower energy transfers. The assignment of the transitions I - III is shown in Fig.~\ref{fig3}; for a
complete discussion of the 0~kbar spectra and assignment of the peaks we refer to Ref.~\onlinecite{Sie05}.
With increasing pressure, the spectra show significant changes, most notably shifts in the position of peaks
I and III. Peak I clearly moves to higher energies with increasing pressure (the position of peak II is
largely unaffected), while peak III moves to lower energy and approaches peak IV, overlapping with it at the
highest pressures.

%

As demonstrated from magnetization,\cite{Wer02} electron spin resonance,\cite{Hil03} and INS
measurements,\cite{Sie05} the low-temperature magnetism in the (Mn$_4$)$_2$ dimer is accurately described as
follows. Each of the Mn$_4$ subunits has a $S$ = 9/2 ground state with an additional zero-field-splitting of
the easy-axis type. Accordingly, the $S$ = 9/2 level is split into five $\pm M_S$ Kramers doublets, which is
well reproduced by the spin Hamiltonian
\begin{equation}
\label{H}
 \hat{H}_{ZFS} = D[\hat{S}^2_z -{1 \over 3}S(S+1)] + B^0_4 \hat{O}^0_4(S),
\end{equation}
where $\hat{O}^0_4(S) = 35 \hat{S}^4_z - 30 S(S+1)\hat{S}^2_z + 25 \hat{S}^2_z +6 S(S+1)$, and the $z$ axis
coincides with the threefold symmetry axis (or $c$ unit-cell axis). In (Mn$_4$)$_2$, two identical Mn$_4$
units are fused together, leading to the following effective spin Hamiltonian for the dimer
\begin{equation}
\label{Hdimer}
 \hat{H} =  \hat{H}_{ZFS,A} + \hat{H}_{ZFS,B} + J \hat{\textbf{S}}_A \cdot \hat{\textbf{S}}_B.
\end{equation}
Here, $\hat{H}_{ZFS,A}$ and $\hat{H}_{ZFS,B}$ are centered on the two Mn$_4$ subunits $A$ and $B$ and have
the form of Eq.~(\ref{H}), and $J>0$ describes the isotropic AFM exchange interaction between the two units.
The effect of $J$ is to split the Mn$_4$ Kramers doublets into two or three components. For instance, the
lowest-lying Mn$_4$ doublet $|M_S=\pm9/2\rangle$ splits into the two dimer doublets
$|M_{S_A}=9/2,M_{S_B}=-9/2\rangle_{s/a}$ and $|M_{S_A}=9/2,M_{S_B}=9/2\rangle_{s/a}$ as indicated in
Fig.~\ref{fig3} (for a detailed classification of the levels, see Ref.~\onlinecite{Sie05}).

\begin{figure}
\includegraphics[scale=0.75]{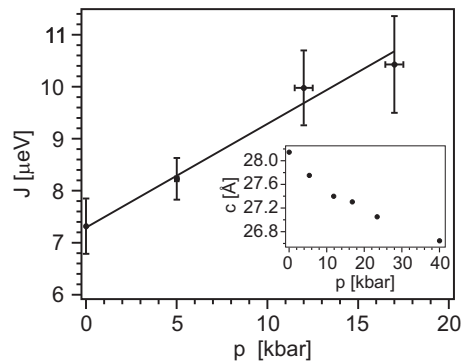}
\caption{\label{fig4} Dependence of $J$ on pressure $p$. The solid line represents the least-square fit
Eq.~(\ref{Jp}). The inset shows the pressure dependence of the unit-cell constant $c$ at room temperature
(error bars are smaller than symbol sizes).}
\end{figure}

From Hamiltonian~(\ref{Hdimer}) INS spectra were calculated as described in Ref.~\onlinecite{Sie05}. The
result of least-square fits to the data, where the parameters $J$, $D$ and $B^0_4$ (and the linewidth) were
chosen as variables, are shown as solid lines in Fig.~\ref{fig2}. The agreement between experimental and
theoretical curves is excellent. Thanks to the many peaks in the data, the fit parameters could be accurately
determined. Within the error bars, a negligibly small pressure dependence of $D$ and $B^0_4$ was detected:
Fitting of the results to a linear dependence yields $D(p)$ = -62.2(2)~$\mu$eV - 0.037(32)~$\mu$eV/kbar
$\times p$ and $B^0_4(p)$ = -7.1(3)$\times$10$^{-3}$~$\mu$eV + 22(56)$\times$10$^{-6}$~$\mu$eV/kbar $\times
p$. The coupling constant $J$ on the other hand is affected strongly, as shown in Fig.~\ref{fig4}. The
best-fit line is
\begin{equation}
\label{Jp}
  J(p) = 7.3(4)~\mu\mathrm{eV} + 0.20(5)~\mu\mathrm{eV}/\mathrm{kbar} \times  p.
\end{equation}

The pressure dependence of the unit-cell parameter $c$ is shown in the inset of Fig.~\ref{fig4}. It
monotonously decreases from 28.168(1)~{\AA} at 0~kbar to 26.5727(9)~{\AA} at 40~kbar in a quasi-linear
fashion. The cell parameter $a$ and the volume $V$ exhibit pressure dependencies similar to $c$, with a
decrease of $a$ from 16.2991(3) to 15.143(1)~{\AA} and of $V$ from 6480.6(1)~{\AA}$^3$ to 5276.8(4)~{\AA}$^3$
in the pressure range 0 - 40~kbar.

The trigonal symmetry axis of the molecular dimers is collinear with the hexagonal axis of the unit cell, and
the decrease of $c$ with pressure directly relates to a decrease of the separation between the Mn$_4$
subunits within the dimer. At 40~kbar $c$ is decreased by a moderate 6\%. Furthermore, the pressure
dependence of the volume $V$ shows a quasi-linear decrease, showing that the pressure range where repulsive
forces between the dimers become significant has not yet been reached. Finally, the bulk modulus $B$ and its
pressure derivative were obtained by fitting a universal equation of state.\cite{Vin87} The resulting small
bulk modulus of $B$ = 11~GPa is consistent with a weak intermolecular bonding. It is hence concluded that the
compression under pressure is achieved mostly through reduction in intermolecular space.

%

The insignificant pressure dependence of $D$ in (Mn$_4$)$_2$ is in clear contrast to the observed 4\%
decrease of $D$ in the SMM Mn$_4$Br, which is very similar to the Mn$_4$ subunits in (Mn$_4$)$_2$, only that
the apical Cl ion is replaced by a Br ion.\cite{Sie04} However, this can be understood by the fact that Mn-Cl
bonds are stronger and less compressible than Mn-Br bonds. Thus the Mn$_4$ subunits of (Mn$_4$)$_2$ are less
distorted under pressure than the Mn$_4$Br cluster, which is consistent with the conclusions of the previous
paragraph.

In contrast, the $J$ value exhibits a large increase from 7.3(4) to 10.4(9)~$\mu$eV in the pressure range 0 -
17~kbar, i.e., an increase of 42\%. Interestingly, this change is of the order of what can be achieved by a
chemical modification, e.g., a solvent exchange: For the cluster studied here, but with the MeCN solvent
molecules replaced by n-hexane molecules, the $J$ value was found to be 10.3(9)~$\mu$eV, or 43\% larger,
while the Cl-Cl distance was decreased by 0.13~{\AA} or 3.4\%, respectively.\cite{Hil03} We conclude that in
(Mn$_4$)$_2$ the Cl-Cl distance between the two Mn$_4$ units decreases by about the same amount of ca. 3.4\%
upon application of 17~kbar. This is further supported by our X-ray experiment, which shows that at 17~kbar
the $c$ axis is reduced by 2.6\%. This represents a lower limit for the compression of the Cl-Cl distance,
assuming a homogeneous compression along $c$. Considering the stiffness of the Mn$_4$ subunits derived above,
the compression of Cl-Cl will be bigger than 2.6\%, and a value of 3.4\% or 0.13~{\AA} at 17~kbar appears
very reasonable.

The functional dependence of $J$ on the Cl-Cl distance $d$ can be expressed as $J(d) = J_0 \exp[-k(d-d_0)]$.
$k$ is a measure of how fast the overlap of the neighboring wave functions decreases with increasing
distance. With $J_0$ and $d_0$ the values at 0~kbar, one obtains $k$ = -2.7(7)~{\AA}$^{-1}$. Recently, the
weak exchange between the subunits in the (Mn$_4$)$_2$ dimer was studied theoretically by means of DFT
calculations.\cite{Par03} The absolute value of $J$ was calculated about a factor of two too large, and for
the dependence on the Cl-Cl distance a slope of $k$ = -2~{\AA}$^{-1}$ was inferred. The latter value should
be reliable as DFT should capture trends better than absolute values. The experiment and the DFT calculation
thus show a nice agreement. For comparison, for the distance dependence of $J$ in iron(III) dimers and
trimers values of $k$ = -6~{\AA}$^{-1}$ and -8~{\AA}$^{-1}$ were found, respectively.\cite{Gor97,Wei97} This
underpins the soft character of the exchange bridge in the (Mn$_4$)$_2$ dimer.

Our results suggest an interesting effect of an applied pressure on the magnetic hysteresis curve. Starting
from negative magnetic fields, (Mn$_4$)$_2$ at ambient pressure undergoes tunneling transitions at fields of
ca. -0.35~T (tunneling from $|9/2,9/2\rangle$ to $|9/2,-9/2\rangle$), at ca. 0.2~T (tunneling from
$|9/2,9/2\rangle$ to $|7/2,-9/2\rangle$), and further transitions at higher fields.\cite{Wer02} Thus, because
of the effect of the inter-dimer coupling $J$ no tunneling transition at zero field is observed, in
remarkable contrast to the behavior in standard SMMs. However, with increasing $J$ the second tunneling
transition $|9/2,9/2\rangle \rightarrow |9/2,-9/2\rangle$ will move to lower and eventually to negative
magnetic fields. This can be inferred from Fig.~\ref{fig3}. In order to fulfill the tunneling condition of
equal energy of both states, the zero-field gap between the states $|9/2,9/2\rangle$ and $|9/2,-9/2\rangle$
has to be overcome by a magnetic field. These two levels, however, approach each other with increasing $J$
and eventually cross at a value of $J_c$ = 14.5~$\mu$eV. At this point, the system will exhibit a tunneling
transition at zero field analogical to standard SMMs. Extrapolating our determined $J(p)$ dependence, this
point is estimated to be reached at ca. 36~kbar, which is well within the bounds of possibility in
magnetization measurements. Thus, it is predicted that as function of pressure the magnetic hysteresis curve
should show strong changes, and that the (Mn$_4$)$_2$ dimer system may be tuned into a condition where it
exhibits a magnetic hysteresis curve with a step at zero field, more similar to that of conventional SMMs.

%

In conclusion, we have studied the response of dimeric SMM (Mn$_4$)$_2$ to hydrostatic pressure, as reflected
in changes of INS and powder X-ray diffraction spectra. The magnetic parameters of the Mn$_4$ subunits were
found to be pressure independent up to 17~kbar, in contrast to findings where changes on the order of 4\%
were observed,\cite{Sie05,Sie05b} but in agreement with the particular chemical situation in this molecule.
On the other hand, the inter-dimer exchange parameter $J$ displayed a pronounced increase by 42\% between
ambient pressure and 17~kbar. The dependence of $J$ on the Cl-Cl-distance between the Mn$_4$ subunits was
estimated and found to be in nice agreement with predictions of DFT calculations, which indicates that DFT
calculations are an appropriate tool to study weak long-distance exchange interactions. From our results we
predict that within experimentally reasonable pressures the magnetic properties of (Mn$_4$)$_2$ will be
drastically affected. At a pressure of about 36~kbar the system should exhibit a magnetic hysteresis curve
more similar to that of a conventional SMM. This tuning possibility could be interesting for potential
applications.

This work demonstrates that the combination of INS and pressure is an excellent tool to study intermolecular
magnetic interactions. Firstly, pronounced changes in the magnetic behavior can be induced, which may be as
large as achieved by chemical variations. Secondly, compared to the application of pressure by chemical
means, hydrostatic pressure can be adjusted continuously which allows for many measurement points.

%

\begin{acknowledgments}
We thank A. J. Church and C. M. Goodway from the ISIS sample environment group for help. Financial support by
EC-RTN-QUEMOLNA, contract n$^\circ$ MRTN-CT-2003-504880, and the Swiss National Science Foundation (NFP 47)
is acknowledged.
\end{acknowledgments}

%

%
\end{document}